\documentclass{aa}
\usepackage{graphicx}
\usepackage{pdfpages}
\usepackage{txfonts}
\usepackage{epsfig}
\usepackage{pdffig}
\usepackage{natbib}
\usepackage{multirow}
\usepackage{lscape}
\usepackage{mathrsfs,amssymb}
\usepackage{amsmath}
\usepackage{lineno}
\usepackage{booktabs}
\usepackage{xcolor}
\newcommand       \Angstrom     {\,{\rm \AA}}

\newcommand       \cm           {\,{\rm cm}}

\newcommand       \erg          {\,{\rm erg}}
\newcommand       \eV           {\,{\rm eV}}

\newcommand       \K            {\,{\rm K}}

\newcommand       \s            {\,{\rm s}}
\newcommand       \sr           {\,{\rm sr}}

\newcommand       \mum          {\,{\rm \mu m}}

\newcommand       \Teff         {T_{\bigstar}}
\newcommand       \Tstar      {T_{\rm eff}}

\newcommand       \simali       {\sim\,}

\newcommand       \EUV         {\langle h\nu\rangle_{\rm abs}}
\newcommand       \Ephoton         {E_{\rm ph}}
\newcommand       \ICN         {I_{\rm CN}}
\newcommand       \ICH         {I_{\rm CH}}
\newcommand       \PCN         {P_{\rm CN}}
\newcommand       \tauabs    {\tau_{\rm abs}}
\newcommand       \Icasc    {I_{\rm casc}}

\begin{document} 


\title{Anharmonic Infrared Emission of
         Cyano-Substituted Polycyclic Aromatic
         Hydrocarbon Molecules:
         Cyanonaphthalenes as a Case Study}

\titlerunning{Anharmonic Emission of Cyanonaphthalenes}
       
\author{Tao~Chen\inst{1,2},
        Kaijun~Li\inst{3,4}, \and
        Aigen~Li\inst{4}
          }
\authorrunning{Chen, Li \& Li}
          
\institute{Xinjiang Astronomical Observatory, 
                Chinese Academy of Sciences,
                150 Science 1-Street, 
                Urumqi, Xinjiang, 
                830011, China\\
\email{chentao@xao.ac.cn}
\and
Division of Theoretical Chemistry and Biology, 
          School of Engineering Sciences in Chemistry, 
          Biotechnology and Health,
          KTH Royal Institute of Technology, 
          SE-100 44 Stockholm, Sweden
\and
Hunan Key Laboratory for Stellar and Interstellar
Physics and School of Physics and Optoelectronics,
Xiangtan University, Hunan 411105, China\\
\email{likj@xtu.edu.cn}
\and
Department of Physics and Astronomy,
             University of Missouri,
             Columbia, MO 65211, USA\\
\email{lia@missouri.edu}
             }
\date{Received February 6, 2026; accepted July 21, 2026}

\abstract
{ The recent radio-spectroscopic detection of nitrile-bearing aromatic
  species, has identified cyano-substituted polycyclic aromatic
  hydrocarbons (cyano-PAHs) as key tracers of nitrogen heterocycles in
  the interstellar medium. Despite this, their infrared (IR)
  identification remains a challenge, requiring high-fidelity
  anharmonic vibrational radiative models that account for dynamic cooling processes in  diverse astrophysical environments.}
{ We aim to provide IR cascade model emission spectra for the anion,
  neutral, and cation states of 1- and 2-cyanonaphthalene in representative astrophysical regions. 
  We investigate how charge states and substitution sites (1- vs. 2-position) 
  modulate the anharmonic vibrational profiles and energy redistribution during the IR cascade process.}
{ We employed B3LYP/N07D in conjunction with the Second-order
  Vibrational Perturbation Theory (VPT2). To
  simulate the emission under astrophysical conditions, we utilized an
  optimized microcanonical sampling algorithm to determine
  the vibrational density of states and generate IR cascade emission spectra
  for different astrophysical environments.}
{ Our results reveal that, compared to neutrals, the intensity of the C$\equiv$N nitrile
  band of anions is dramatically enhanced by an order
  of magnitude, while the C$\equiv$N band intensity of cations is
  somewhat weaker than that of neutrals. We show that 1- and
  2-cyanonaphthalene exhibit distinct isomeric signatures:
  1-cyanonaphthalene induces a complex ``red-wing'' structure in the
  3.3$\mum$ C--H stretch due to peri-hydrogen interactions, whereas
  2-cyanonaphthalene displays a more symmetric profile. We
  evaluated the fractional energy emitted via the 
  C$\equiv$N stretch relative to the total IR cascade intrinsic
  emissivity for neutral, cationic, and anionic 1- and
  2-cyanonaphthalene in different astrophysical environments, and
  demonstrated how they, combined with the observationally detected
  C$\equiv$N stretch flux, can be used to quantitatively probe their presences
  in space.}
{ By bridging the gap between static laboratory benchmarks and dynamic
  interstellar emission, this work provides accurate anharmonic
  cascade emission spectra for cyanonaphthalenes. The spectral
  fingerprints for different charge states and isomers offer essential
  tools for interpreting high-resolution observations from the
  \textit{James Webb Space Telescope}.}
\maketitle

\section{Introduction}
Polycyclic Aromatic Hydrocarbon (PAH) molecules
are widely recognized as a fundamental component
of the interstellar medium (ISM), accounting for
a substantial fraction of the cosmic carbon reservoir
\citep{Tielens2008, Li2020}.
Since the formulation of the PAH hypothesis,
the prominent infrared (IR) emission bands observed
at 3.3, 6.2, 7.7, 8.6, and 11.3$\mum$ have been
attributed to the collective vibrational modes
of these large carbonaceous species
\citep{Leger1984, Allamandola1985, Allamandola2021}.
These features, commonly referred to
as aromatic IR bands (AIBs),
are now routinely observed across a wide range
of astrophysical environments,
from the Galactic diffuse ISM,
photodissociation regions, and circumstellar
disks to distant galaxies \citep{Li2020}.
Despite this apparent ubiquity, the definitive identification
of individual PAH molecules through IR spectroscopy
remains a long-standing challenge.

A major breakthrough occurred with the detection of benzonitrile
($\text{C}_6\text{H}_5\text{CN}$) \citep{Kwon2003, Rajasekhar2022} in the cold molecular cloud TMC-1
using radio spectroscopy \citep{McGuire2018}. This was followed by the
landmark identification of the first individual cyano-substituted
PAHs: 1- and 2-cyanonaphthalenes \citep{McGuire2021}. The polar cyano
($-$CN) group not only enables the detection of these species via
rotational transitions but also serves as a chemical tracer for the
nitrogen heterocyclic chemistry in the ISM \citep{Burkhardt2021}.
These discoveries suggest that cyano-PAHs may be widespread
in the ISM. Indeed, in subsequent years, several additional
cyano-PAH species have been detected through their rotational
spectra, including 2-cyanoindene (C$_9$H$_7$CN; \citealt{Sita2022}),
1- and 5-cyanoacenaphthylene
(C$_{12}$H$_{7}$CN; \citealt{Cernicharo2024}),
1-, 2- and 4-cyanopyrene (C$_{16}$H$_9$CN;
\citealt{Wenzel2024a,Wenzel2024b}),
and cyanocoronene (C$_{24}$H$_{11}$CN; \citealt{Wenzel2025}).

However, none of these aforementioned, rotationally
identified specific aromatic molecules has yet been
detected in the IR. The difficulty of identifying specific
PAHs through their IR spectra arises primarily from
the intrinsic similarity of their vibrational spectra.
Nevertheless, the IR spectra of different PAH molecules
are readily distinguishable in the laboratory
from one species to another.
Therefore, in principle, IR spectroscopy could
provide individual identification if the molecule 
is sufficiently abundant and the astronomical
detectors are sufficiently sensitive.
In our opinion, the (so far) nondetection of
individual PAH molecules in the IR is partly
attributed to the lack of knowledge of their
exact IR spectra and their phys-chemical properties. 
Due to the inherent challenges in molecular isolation, 
experimental studies on cyano-PAHs remain scarce in 
the literature, particularly for larger species and 
those in charged states. To date, experimental 
investigations have been primarily constrained to 
smaller species in cationic and anionic 
forms \citep{Dixon2015, Kirnosov2017, Gulania2020, Liu2022, Palotas2024, Stockett2025,Jiang2026,Zhang2026}.
Also, unlike the trending
``GrandPAH'' hypothesis which suggests
only a limited number of different PAHs
are present in the ISM (which implies that the abundance
of an individual molecule might not be very low; 
\citealt{Andrews2015}), the abundances of those
aforementioned, rotationally identified individual
molecules are rather low
(e.g., see \citealt{McGuire2021};
\citealt{Burkhardt2021}; \citealt{Sita2022};
\citealt{Wenzel2024a, Wenzel2024b, Wenzel2025};
\citealt{Cernicharo2021, Cernicharo2024};
\citealt{Cabezas2025}).
Thus, their IR signals are expected to be faint.
While it may be challenging
for the {\it Infrared Space Observatory} (ISO)
and even the {\it Spitzer Space Telescope}
to detect such weak signals,
with the advent of 
the {\it James Webb Space Telescope} (JWST),
this may become possible
due to its unprecedented sensitivity
and wavelength coverage
(particularly in the near-IR
unexplored by Spitzer).

With an aim to offer spectroscopic guidance
for JWST to search for specific PAH molecules,
\citep{Li2024a, Li2024b} calculated the IR emission
spectra of 1- and 2-cyanonaphthalenes and
indene expected in different astrophysical
environments. However, their model emission
spectra were computed by making use of
the frequencies and intensities of the vibrational
transitions computed from harmonic approximations,
which inherently neglect anharmonic effects
such as frequency shifts, intensity redistribution,
and mode coupling (e.g., see \citealt{Mackie2015, Mackie2018, Chen2018aa}).
More recent investigations have demonstrated that,
for cyano-PAHs, the C$\equiv$N stretching feature
near 4.4$\mum$ is not a simple isolated band
but rather a structured profile shaped by anharmonic
couplings and resonance polyads
\citep{Boersma2023, Esposito2024a, Esposito2024b}.
These findings emphasize that a fully anharmonic
treatment is essential for accurately modeling
the IR emission spectra of specific PAHs
in astrophysical environments.

In astrophysical environments, PAHs are typically
excited by single ultraviolet (UV) photons and
subsequently relax through a sequence of
IR emission \citep{Draine2001}, a process commonly described
as an anharmonic cascade \citep{Stockett2025, Xu2024a, Xu2024b}.
Modeling this cascade requires not only accurate
anharmonic vibrational data but also a robust
statistical description of the vibrational energy
distribution. Techniques such as microcanonical
sampling, including the Wang–Landau (WL)
algorithm \citep{Wang2001},
provide an efficient framework
for constructing the vibrational density of states
and simulating energy-resolved emission processes.

In this work, we investigate the anharmonic vibrational
emission spectra of 1- and 2-cyanonaphthalenes
across three charge states (neutral, cation, and anion)
in a number of representative astrophysical environments.
By combining high-level anharmonic vibrational calculations
(\S\ref{sec:method}) with the Wang–Landau microcanonical
sampling and cascade emission modeling,
we compute the anharmonic  absorption spectra
(\S\ref{sec:anharm}) and physically realistic IR emission
spectra (\S\ref{sec:emsn}) of neutral, cationic and anionc
1- and 2-cyanonaphthalenes. Our results are benchmarked
against recent cryogenic gas-phase experiments,
ensuring both theoretical accuracy and astrophysical relevance
(\S\ref{sec:astro}). The resulting model emission spectra provide
critical guidance for searching for and identifying cyanonaphthalenes
in the JWST era (\S\ref{sec:astro}).

\section{Computational Methods}\label{sec:method}
The equilibrium geometries and vibrational
properties 1- and 2-cyanonaphthalenes;
(hereafter 1-CNN and 2-CNN) across three
charge states (neutral, cation, and anion)
were investigated using the Gaussian 16
software package \citep{g16}.
To ensure high-precision potential energy surfaces,
a two-step optimization strategy was employed.
Initial structural searches were performed
at the B3LYP/3-21G level \citep{Lee1988},
followed by refinement using the B3LYP functional
in conjunction with the N07D basis set \citep{Barone2008},
which is specifically optimized for the spectroscopic
properties of medium-sized organic molecules. 

Anharmonic vibrational frequencies
and intensities were computed via Second-order
Vibrational Perturbation Theory
\citep[VPT2;][]{Bloino2012}.
To accurately treat the complex resonance patterns
characteristic of polycyclic aromatic species,
we adopted a Generalized VPT2 (GVPT2) protocol \citep{Bloino2015}.
Near-degeneracies, such as Fermi resonances,
were identified through a deperturbed model and
subsequently handled variationally to ensure
the reliability of spectral profiles in the critical
C--H stretching ($\simali$3.3$\mum$) and
C$\equiv$N stretching ($\simali$4.4$\mum$) regions.

To simulate the IR emission under interstellar conditions,
we utilized an optimized WL microcanonical sampling
algorithm \citep{Wang2001} to determine
the vibrational density of states (DOS)
and energy-dependent emission profiles.
The internal energy domain was sampled from $0$ to $15\eV$
with a resolution of $0.001\eV$,
ensuring a converged representation of
the vibrational phase space up to the Lyman limit.
During the random walk in the quantum number space,
an incremental update scheme was implemented to
evaluate the VPT2 Hamiltonian,
where the energy and effective frequencies
were updated based on the local changes in
vibrational quanta rather than full matrix recalculation.
This approach significantly accelerated the accumulation
of spectral power distributions as a function of internal energy.

To simulate the physically realistic IR emission of 1- and 2-CNN 
under interstellar conditions, we employ a microcanonical approach 
to model the relaxation process following single-photon UV 
excitation. The simulation leverages an optimized Wang–Landau (WL) 
sampling algorithm \citep{Wang2001, Chen2018aa, Mackie2018} to 
determine the vibrational density of states (DOS) and 
energy-dependent emission profiles through three hierarchical stages:

\begin{enumerate}
\item DOS Calculations: The WL algorithm is utilized to sample 
the vibrational phase space defined by the anharmonic GVPT2 
Hamiltonian. By iteratively updating the density of states $g(E)$ 
and an auxiliary histogram to ensure a flat sampling distribution 
across the internal energy range (up to $15\eV$), we obtain the 
converged microcanonical DOS for all charge states of 1- and 2-CNN.

\item Accumulated Emission Spectra: During the random walk, 
the instantaneous transition intensities---encompassing fundamentals, 
overtones, and combination bands---are accumulated into designated 
energy bins ($\Delta E$). This yields the \textit{accumulated spectra} 
$S(E,\nu)$, which map the spectral power distribution at specific 
internal energies $E$, inherently incorporating the anharmonic 
redshifts and broadening driven by increased vibrational quanta.

\item Cascade Emission Spectra: The final IR cascade spectrum, 
representing the total emitted energy per wavenumber ($\eV\cm$) during 
a complete starlight photon cooling event, is computed by integrating 
the normalized accumulated emission over the entire cooling trajectory:
\begin{equation}\label{eq:Icasc}
\Icasc(\nu) = \int_{0}^{\Ephoton} \frac{S(E,\nu)}{P(E)} dE ~~,
\end{equation}
where $\Ephoton$ is the energy of the photon
absorbed by the molecule,
and $P(E) = \int S(E,\nu) d\nu$ is the total radiative power at 
internal energy $E$. This statistical integration explicitly accounts 
for the fact that the initial vibrationally hot states 
dominate the anharmonic line broadening and ``red wings,'' whereas 
the molecule spends the majority of its cooling lifetime in lower-energy, 
``narrower'' vibrational states as it relaxes toward the ground state.
\end{enumerate}

\section{Anharmonic Absorpion Spectra}\label{sec:anharm}
The theoretical treatment employed B3LYP/N07D with VPT2
to account for anharmonicity, including the treatment of
Fermi resonances. The \texttt{N07D} basis set is widely 
recommended for anharmonic vibrational calculations, as 
it provides an optimal compromise between chemical 
accuracy and computational overhead \citep{Barone2008}. 
For complex organic molecules and PAHs of interstellar interest, 
the B3LYP/N07D model chemistry has demonstrated remarkable reliability 
in capturing vibrational manifold \citep{Chen2018apjs},
even when benchmarked against significantly
more expensive computational schemes
\citep{Esposito2025, Xu2024b}. 
A prominent example is the double-hybrid \textit{rev}-DSDPBEP86 method 
combined with the larger \texttt{jun-cc-pVTZ} basis set (utilizing 
B3LYP/N07D for the cubic and semidiagonal quartic force constants, 
hereafter abbreviated as rDSD-TZ+B3LYP). Despite its higher 
theoretical level, the rDSD-TZ+B3LYP method does not 
necessarily yield superior results for specific nitrile species. 
In particular, for 2-cyanoindene, the mean absolute error (MAE) 
between theory and experiment is 6.5 or 4.1$\cm^{-1}$ 
(including or excluding the CN stretch) for the B3LYP/N07D method, 
compared to 6.8 or 6.5$\cm^{-1}$ for the rDSD-TZ+B3LYP approach.
Although the rDSD-TZ+B3LYP hybrid does not yield a 
uniform improvement across the complete vibrational 
spectrum of cyano-substituted species, it offers 
a pronounced advantage in reproducing the CN-stretching 
feature of neutral CN-PAHs \citep{Stockett2025}.

\begin{figure*}
\centering
\includegraphics[width=\textwidth]{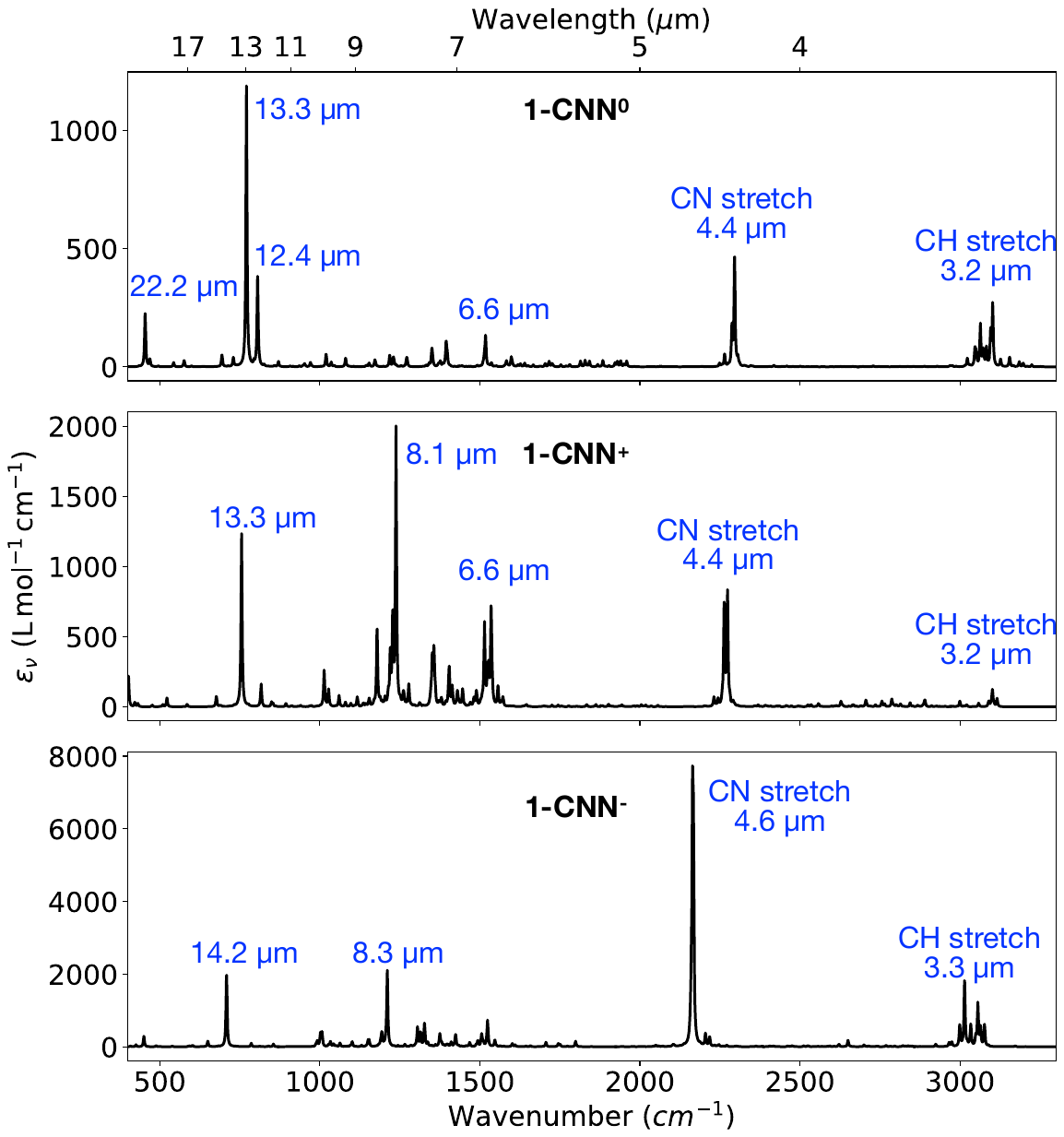}
\caption{DFT-computed anharmonic absorption
spectra of 1-cyanonaphthalene in its neutral
($1\text{-CNN}^0$, top panel),
cation ($1\text{-CNN}^+$, middle panel),
and anion ($1\text{-CNN}^-$, bottom panel) charge states.
The calculated vibrational transitions are convolved
with a Lorentzian profile using a FWHM of 4$\cm^{-1}$
to simulate realistic spectral profiles.
Numerical labels near the major peaks represent
their respective wavelengths. 
Key vibrational bands, including the CN stretch
at $\simali$4.4--4.6$\mum$
and the CH stretch at $\simali$3.2--3.3$\mum$,
are explicitly highlighted.
}
\label{fig:fig1}
\end{figure*}

\begin{figure*}
\centering
\includegraphics[width=\textwidth]{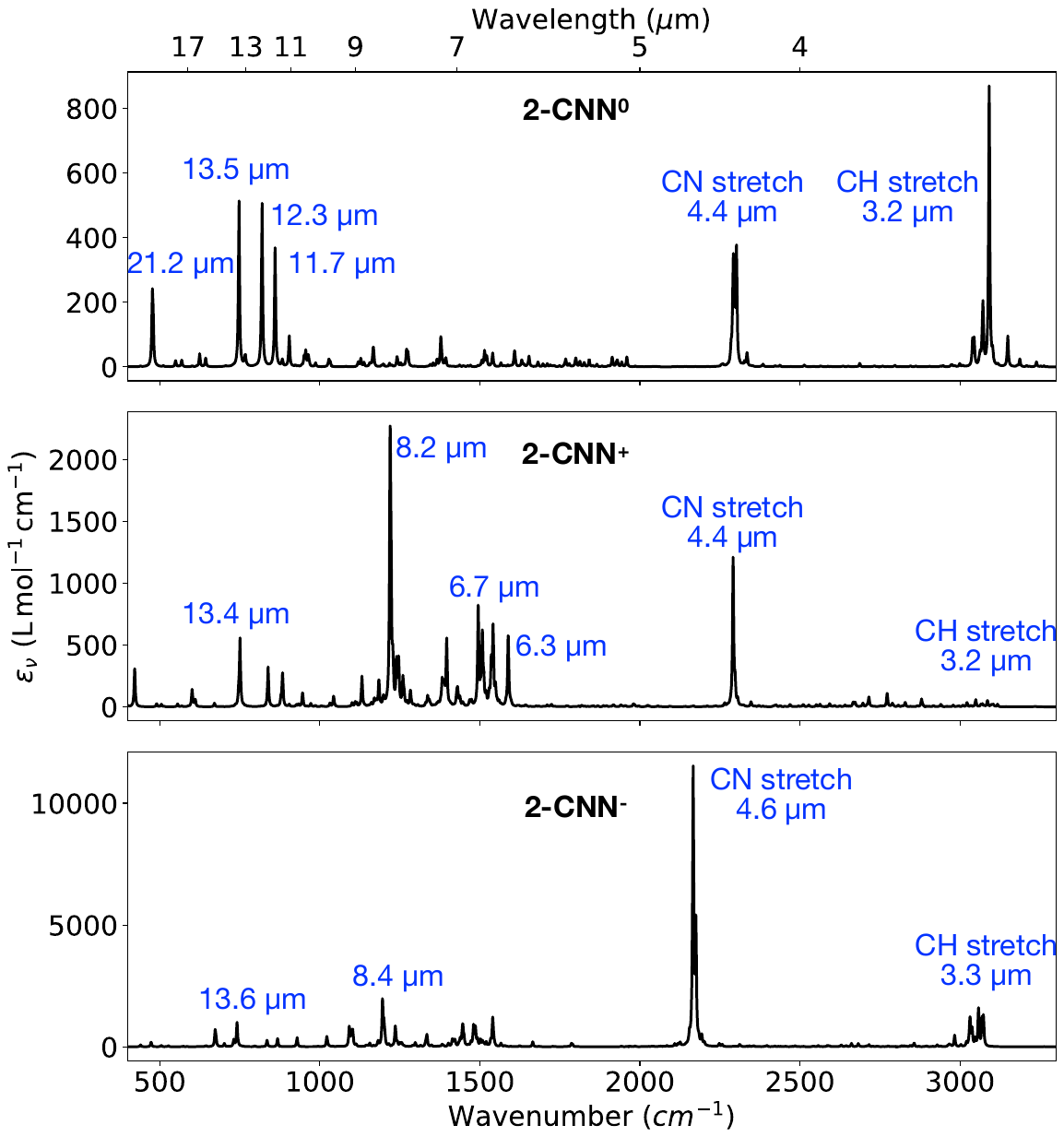}
\caption{Same as Figure~\ref{fig:fig2}
but for 2-cyanonaphthalene in its neutral
($2\text{-CNN}^0$, top panel),
cation ($2\text{-CNN}^+$, middle panel),
and anion ($2\text{-CNN}^-$, bottom panel) states.
}
\label{fig:fig2}
\end{figure*}

We show in Figures~\ref{fig:fig1} and \ref{fig:fig2}
the anharmonic absorption spectra of 1-CNN and
2-CNN in their neutral, cationic, and anionic charge states.
All spectra were generated from anharmonic vibrational
calculations and subsequently convoluted
using a Lorentzian profile with
a full-width at half-maximum (FWHM)
of 4$\cm^{-1}$ to facilitate comparison with
experimentally accessible spectral resolutions
(see Figure~\ref{fig:fig5}).

For both isomers, several characteristic spectral regions
can be clearly identified. The C$\equiv$N stretching mode
appears as a prominent and relatively isolated feature
in the 4.4--4.6$\mum$ region, while the aromatic C--H
stretching modes are clustered near 3.2--3.3$\mum$.
In addition, a series of bands associated with C--C
stretching and C--H in-plane and out-of-plane
bending modes populate the 6--15$\mum$ region.
Despite the overall similarity in spectral patterns,
significant variations in both band positions and
intensities are observed across different charge states.

As shown in Figure~\ref{fig:fig2}, the anharmonic
spectrum computed for neutral 1-CNN exhibits
a moderately strong C$\equiv$N stretching band
near 4.4$\mum$ and relatively well-defined features
in the mid-IR region, including bands around 6.6, 12.4,
13.3 and 22.2$\mum$. Upon ionization to the cation,
the spectrum becomes more congested, with enhanced
intensities in the 6--9$\mum$ region and a noticeable
redistribution of oscillator strengths.
For anionic 1-CNN, most notable is the strongly
enhanced C$\equiv$N stretching band.
This band shifts to longer wavelength
(around 4.6$\mum$) and exhibits substantial
intensity amplification compared to its neutral
and cationic counterparts. This behavior can be traced
back to fundamental changes in their electronic structure.
In particular, the addition of an extra electron in the anion
leads to the occupation of anti-bonding $\pi^{\star}$ orbitals,
which reduces the effective bond order of key functional
groups such as the C$\equiv$N and C--H bonds.
This weakening of chemical bonds results in
a decrease in the associated force constants $k$,
thereby lowering the vibrational frequencies $\nu$.
Within the harmonic approximation,
the vibrational frequency is given by
\begin{equation}
\nu = \frac{1}{2\pi} \sqrt{\frac{k}{\mu}} ~~,
\end{equation}
where $\mu$ is the reduced mass of the oscillator.
Consequently, a reduction in $k$ leads to a decrease
in $\nu$, corresponding to a systematic redshift
in the vibrational spectrum. This effect is particularly
pronounced for the C$\equiv$N stretching mode,
which is highly sensitive to changes in the electronic
distribution due to its partially localized character.

A similar trend is observed for 2-CNN
(see Figure~\ref{fig:fig3}),
although subtle structural differences lead to
distinguishable spectral variations.
The neutral 2-CNN spectrum shows a slightly
different distribution of mid-IR bands,
particularly in the 11--14$\mum$ region.
The cation spectrum again exhibits increased
spectral complexity and intensity redistribution,
while the anionic 2-CNN displays a pronounced
and sharply defined C$\equiv$N stretching band
near 4.6$\mum$, similar to anionic 1-CNN.
However, differences in relative band intensities
and peak positions between the two isomers
indicate that isomer-specific structural effects
remain important, even in the presence of
strong charge-induced perturbations.

Taken together, these results demonstrate
that both charging and the substitution
pattern significantly influence the anharmonic
vibrational spectra of cyano-PAHs.
In particular, the strong and charge-sensitive
C$\equiv$N stretching mode provides a potentially
powerful diagnostic feature for identifying such species
in astrophysical environments.
The systematic redshift and intensity enhancement
observed for anionic species suggest that negatively
charged cyano-PAHs may contribute distinct signatures
to the interstellar IR emission spectrum.

We note that, while the anharmonic spectra
computed from DFT provide essential information
on band positions, intensities, and mode coupling,
they cannot fully capture the spectral profiles observed
in astrophysical environments. A more realistic description
requires the incorporation of vibrational energy redistribution
and time-dependent emission processes, typically modeled
within an anharmonic cascade framework.
In the following we will extend our analysis
by combining the present anharmonic data
with microcanonical sampling and cascade emission
simulations to generate emission spectra that are
directly comparable to astronomical observations.

\section{Anharmonic Cascade Emission Spectra}\label{sec:emsn}
The anharmonic cascade spectra of 1- and 2-CNN
in three different charge states (anion, neutral, and cation)
are simulated to predict their emission signatures
in various astrophysical environments.
The cascade spectrum accounts for the complete
energy relaxation process of a molecule following
the absorption of a single UV photon of energy
$\Ephoton$ (see eq.\,\ref{eq:Icasc}).
We approximate $\Ephoton$ by $\EUV$,
the mean energy of the photons absorbed by
the molecule in an astrophysical environment:
\begin{equation}
\EUV = \tauabs^{-1}\,\int_{0}^{\infty}
C_{\rm abs}(\lambda)\,4\pi J_\lambda^{\star}\,d\lambda ~~,
\end{equation}
where $C_{\rm abs}(\lambda)$ is the absorption
cross section at wavelength $\lambda$ of the molecule,
$J_\lambda^{\star}$ is the starlight intensity
(in units of ${\rm erg}\s^{-1}\cm^{-3}\sr^{-1}$)
in the astrophysical region, and $\tauabs$
is the photon absorption time scale:
\begin{equation}
\tauabs^{-1} = \int_{0}^{\infty}
\frac{C_{\rm abs}(\lambda)\,4\pi J_\lambda^{\star}}  
{hc/\lambda}\,d\lambda ~~,
\end{equation}
where $h$ is the Planck constant,
and $c$ is the speed of light.

Apparently, the mean photon energy $\EUV$,
the photon absorption rate $\tauabs^{-1}$,
and the resulting cascade emission spectrum
of the molecule depend on the starlight spectrum
and intensity. Following Li et al.\ (2004a,\,b),
we consider four representative astrophysical
environments characterized by different starlight
spectra. We first consider the diffuse ISM 
excited by the solar neighborhood interstellar radiation
field of Mathis et al.\ (1983; hereafter MMP83).
For simplicity, we do not distinguish the UV absorption
cross sections of 1- and 2-CNN neutrals, cations and
anions (see Figure~2 of Li et al.\ 2024).
We derive the mean photon energy to be
$\EUV\approx8.6\eV$ and the photon absorption
time scale to be $\tauabs\approx3.71\times10^7\s$
for all these species when exposed to the MMP83
radiation field. If the starlight intensity is enhanced
by a factor of $U$, where $U$\,=\,1 corresponds to
the MMP83 radiation field, the mean photon energy
will remain unchanged, while the photon absorption
time scale will be shortened by a factor of $U$
(i.e., $\tauabs \propto U^{-1}$).

We have also simulated the cascade emission of
cyanonaphthalenes exposed to stars with three
different effective temperatures:
$\Teff$\,=\,40,000, 22,000, 8,000$\K$.
The starlight spectra are approximated by
the Kurucz model atmospheric spectra,
and their intensitis are measured in terms
of the MMP83 field:
%
\begin{equation}
U = \frac{\int_{1\mu {\rm m}}^{912{\rm \Angstrom}}
               4\pi J_\lambda^\star(\Tstar)\,d\lambda}
       {\int_{1\mu {\rm m}}^{912{\rm \Angstrom}}
               4\pi J_{\rm ISRF}(\lambda)\,d\lambda} ~~,
\end{equation}
where $J_{\rm ISRF}(\lambda)$ is 
the MMP83 radiation intensity, and
$J_\lambda^\star(\Tstar)$ is the intensity
of starlight approximated by the Kurucz model
atmospheric spectrum.
With $\Teff$\,=\,40,000$\K$ (like O6V stars)
and $U=10,000$, such a starlight spectrum
and intensity resemble that of the Orion Bar
photodissociation region
and the M17 star-forming region.
We derive $\EUV\approx9.8\eV$ and 
$\tauabs\approx732\s$.
With $\Teff$\,=\,22,000$\K$ (like B1.5V  stars)
and $U=1,000$ like the reflection nebula
NGC\,2023, we obtain $\EUV\approx9.3\eV$
and $\tauabs\approx8213\s$.
For $\Teff$\,=\,8,000$\K$
(like A5V stars) and $U$\,=\,10$^5$
like the Red Rectangle protoplanetary nebula
(but also see Witt et al.\ 2009),
we determine $\EUV\approx5.7\eV$ and 
$\tauabs\approx816\s$.
Table~\ref{tab:integrated_energy_ratio} lists
the mean photon energies $\EUV$ and the photon
absorption time scales $\tauabs$
for these four radiation fields.
Again, as long as the molecule is
in the single photon heating regime
(i.e., the photon absorption time scale
is much longer than the cooling time scale),
the mean photon energy will remain
independent of $U$, while the photon
absorption will be $U$-times more frequent
so that $\tauabs$ will become $\tauabs/U$
(see Draine \& Li 2001).

\begin{figure*}
\centering
\includegraphics[width=\textwidth]{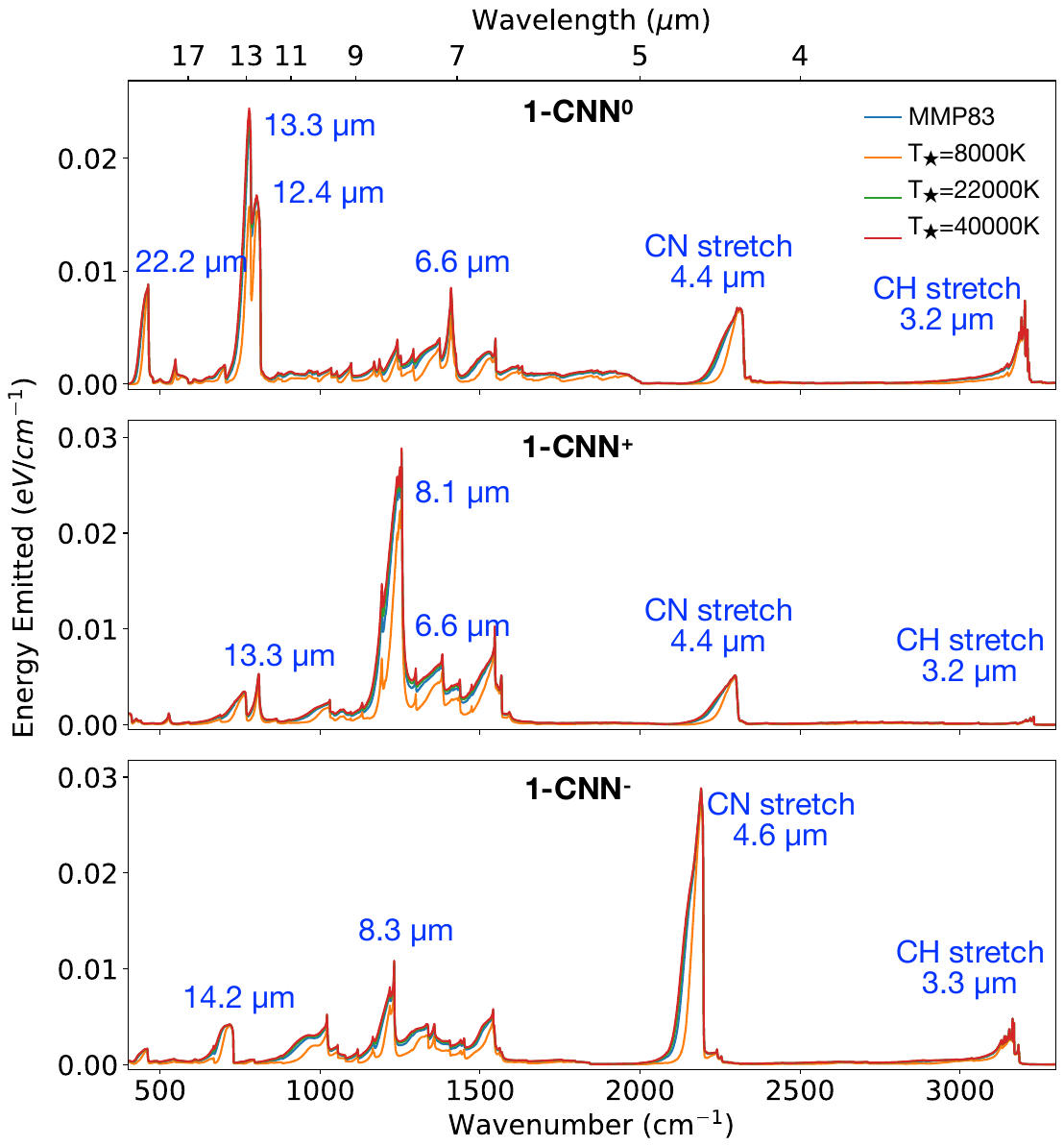}
\caption{Simulated cascade IR emission spectra
of 1-CNN  in its cation (top), neutral (middle),
and anion (bottom) charge states.
Each panel displays the emission profiles
expected in (i) the diffuse ISM (orange)
excited by the MMP83 radiation field
corresponding to a mean UV excitation energy
of $\EUV\approx8.6\eV$,
(ii) the Red Rectangle protoplanetary nebula
(blue) excited by an A5V star of $\Teff$\,=\,8,000$\K$
corresponding to $\EUV\approx5.7\eV$,
(iii) the reflection nebula NGC\,2023 (green)
excited by a B1.5V star of $\Teff$\,=\,22,000$\K$
corresponding to $\EUV\approx9.3\eV$, and
(iv) the Orion Bar photodissociation region
or the M17 star-forming region (red)
excited an O6V star of $\Teff$\,=\,40,000$\K$
corresponding to $\EUV\approx9.8\eV$.
The emission is measured in terms of 
the emitted energy per wavenumber (eV$\cm$)
during a starlight photon absorption event.
Note the enhancement in the total integrated
band intensities along with the pronounced anharmonic
broadening and redshifted ``red wings'' as the
excitation photon energy $\EUV$ increases, which acts
to damp the growth of nominal peak heights for
high-frequency features such as the C--H
and C$\equiv$N stretches.}
\label{fig:fig3}
\end{figure*}

\begin{figure*}
\centering
\includegraphics[width=\linewidth]{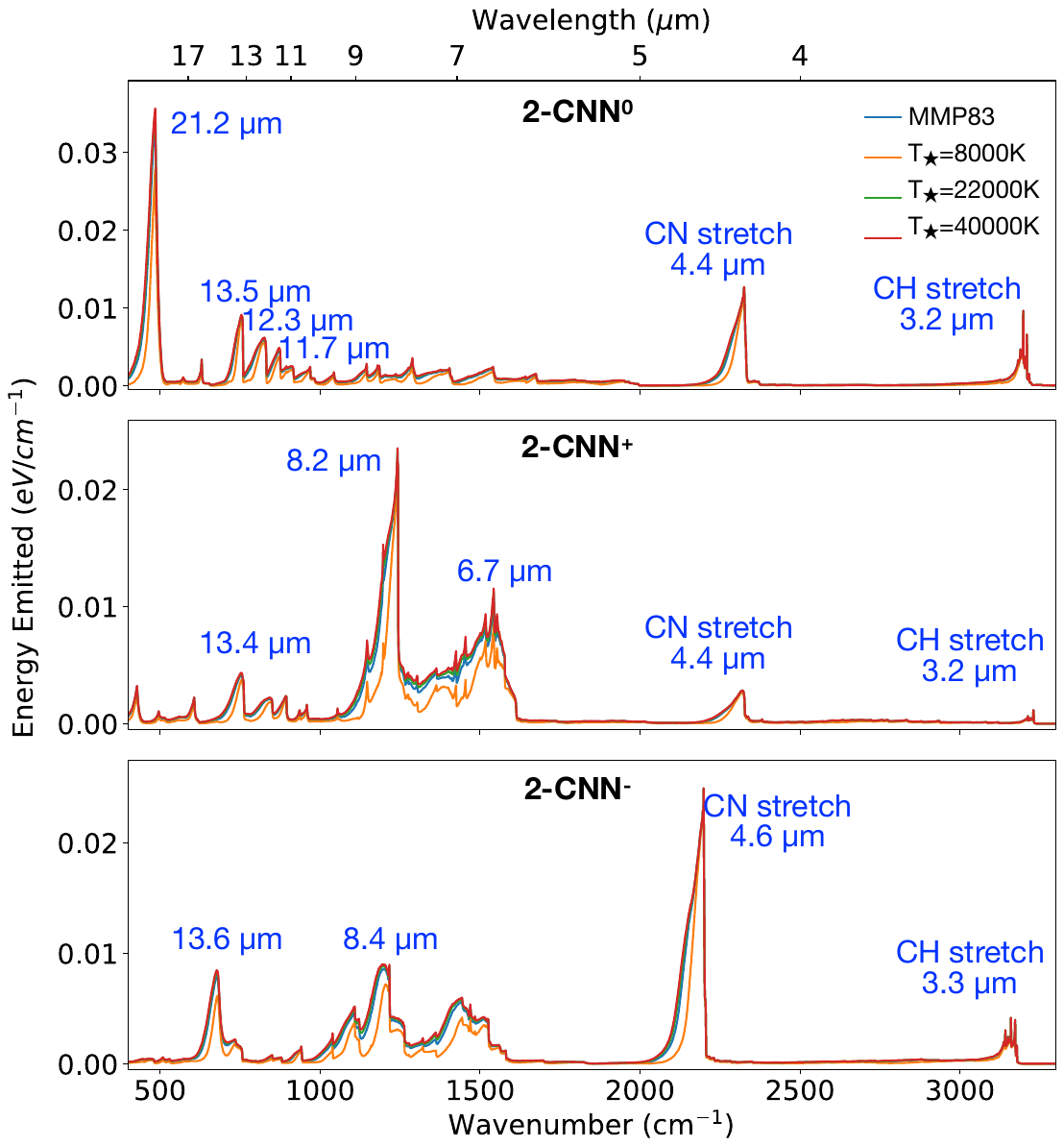}
\caption{
Same as Figure~\ref{fig:fig4} but for 2-CNN.
Compared to 1-CNN, 2-CNN displays a distinct
redistribution of intensities in the skeletal and
nitrile stretching regions.
}
\label{fig:fig4}
\end{figure*}

We show in Figures~\ref{fig:fig3} and \ref{fig:fig4}
the cascade emission spectra respectively
for 1- and 2-CNN neutrals, cations and anions
following UV photon absorption,
computed based on the microcanonical emission
model implemented in our optimized workflow.
For a given initial excitation energy $\EUV$
represented by the mean energy of
the photons absorbed by cyanonaphthalenes
in a specific astrophysical environment
(e.g., $\simali$5.7$\eV$ for the Red Rectangle,
$\simali$8.6$\eV$ for the diffuse ISM,
$\simali$9.3$\eV$ for the reflection nebula NGC\,7023,
$\simali$9.8$\eV$ for the Orion Bar and the M17
star-forming region), the total emitted energy
per unit wavenumber $\Icasc(\nu)$ is obtained
by integrating the energy-dependent emission
profiles $S(E,\nu)$ from the peak excitation
down to the vibrational ground state
(see eq.\,\ref{eq:Icasc}),
with an energy step size of $0.001\eV$.

By normalizing the emission at each energy step
by the total power, the resulting spectrum
represents the amount of energy radiated
per wavenumber (eV$\cm$) throughout
the entire cooling process. This approach naturally
incorporates the anharmonic redshift and line
broadening that occur when the molecule is
``vibrationally hot'' immediately after UV absorption.

For both isomers,
as shown in Figures~\ref{fig:fig3} and \ref{fig:fig4},
the cascade spectra exhibit several prominent
and recurring features across all charge states.
The C$\equiv$N stretching mode
remains a dominant and well-isolated band in
the $\simali$4.4--4.6$\mum$ region,
while the aromatic C--H stretching modes
appear near $\simali$3.2--3.3$\mum$.
In addition, strong emission features are observed
in the 6--9$\mum$ region, associated with C--C
stretching and C--H in-plane bending modes,
as well as in the 11--14$\mum$ region,
corresponding to C--H out-of-plane bending vibrations.
Notably, the cascade spectra display broader and more
structured profiles compared to the computed spectra at 0K,
reflecting the redistribution of vibrational energy
during the emission process.

For 1-CNN, the neutral species shows a relatively
distributed emission pattern, with significant
contributions from bands at 6.6, 12.4, 13.3 and
22.2$\mum$, alongside the characteristic
C$\equiv$N stretching feature
at $\simali$4.4$\mum$ (see Figure~\ref{fig:fig3}).
Upon ionization, the cationic spectrum becomes more
concentrated in the $\simali$6--9$\mum$ region, with
enhanced intensity around 8.1$\mum$ and a noticeable
redistribution of the emission. This spectral
evolution upon ionization---specifically the dramatic
enhancement of the 6--9$\mum$ features relative to
the 3.3$\mum$ band---is a well-established
characteristic common to generic PAHs,
reflecting the intrinsic behavior of C--C stretching
modes in ionized planar aromatics rather than
an effect exclusive to CN-substitution
\citep{Allamandola1999,Hudgins2005}.
In contrast, the anionic spectrum is dominated by a very
strong and narrow C$\equiv$N stretching band near
4.6$\mum$, accompanied by a redshifted C--H stretching
feature around 3.3$\mum$. The relative intensity
variations among different charge states primarily reflect 
intrinsic modifications in the derivatives of the molecular 
dipole moments for specific vibrational modes upon charge 
transfer, as typically mirrored in their counterpart absorption 
spectra, rather than a preferential funneling of starlight 
energy during the emission cascade.

Figure~\ref{fig:fig4} shows that a similar overall behavior
is seen for 2-CNN, although distinct differences arise from
its structural isomerism. The neutral 2-CNN spectrum
shows prominent features at $\simali$12.3--13.5$\mum$
and a strong long-wavelength band near 21.2$\mum$,
which is less pronounced in 1-CNN. The cation again
exhibits enhanced emission at $\simali$6--9$\mum$,
particularly around 8.2$\mum$, while maintaining
a C$\equiv$N stretching feature near 4.4$\mum$.
As in the case of 1-CNN, the anionic 2-CNN spectrum
is dominated by an intense C$\equiv$N stretching band
near 4.6$\mum$ and a redshifted C--H stretching mode
near 3.3$\mum$. However, differences in the mid-R region,
such as the relative intensities of the 8.4 and 13.6$\mum$
bands, highlight the sensitivity of cascade emission
to molecular structure.

An important feature of the cascade spectra
is their dependence on the excitation energy $\EUV$.
As shown in Figures~\ref{fig:fig3} and \ref{fig:fig4},
increasing excitation energy leads to a gradual
enhancement in the total integrated energy of
higher-frequency emission bands---predominantly
accommodated within their extended red wings rather than
causing an increase in their nominal peak heights---while
the lower-frequency bands grow universally in height.
This behavior is governed by the higher initial
microcanonical temperature attained upon
absorbing harder photons,
coupled with the cumulative emission
over a longer cooling cascade history
during which low-frequency modes remain
actively excited down to low temperatures.
Unlike the computed aborption spectra,
which represent transitions from the ground
vibrational state, cascade spectra inherently
encode the full vibrational relaxation pathway,
including anharmonic coupling
and mode competition effects.

Finally, it is imperative to emphasize that
the spectral broadening observed
in our simulated cascade spectra is not
an {\it ad hoc} empirical adjustment
but a physically intrinsic manifestation
of the anharmonic cooling process.
Unlike the computed absorption spectra
shown in Figures~\ref{fig:fig1} and \ref{fig:fig2}
that rely on artificial Gaussian or Lorentzian smoothing,
the line widths and asymmetric profiles presented here
arise naturally from the microcanonical sampling
of the anharmonic vibrations.
As the molecule cascades from high internal energies,
the large population of excited vibrational states
and the associated strength of anharmonic couplings
($\chi_{ij}$) lead to the characteristic ``red wings''
and thermal broadening. This captures the real-time
spectral evolution as the molecule cools,
providing a high-fidelity representation of
the emission physics in the ISM.

\section{Astrophysical Implications}\label{sec:astro}
Compared to the DFT-computed anharmonic
spectra presented in \S2, the cascade emission
spectra exhibit broader features, altered intensity
distributions, and energy-dependent spectral evolution,
all of which are essential for interpreting
high-resolution astronomical observations,
such as those obtained with JWST.
In particular, the strong and charge-sensitive
C$\equiv$N stretching band provides
a promising diagnostic for identifying
cyano-substituted PAHs and
constraining their ionization balance
in different astrophysical environments,
as it has been demonstrated in \S3 that charge
state plays a critical role in shaping the emission
characteristics, particularly for the C$\equiv$N
stretching mode, which exhibits a clear and
systematic shift from $\simali$4.4$\mum$
in neutral and cationic species to
$\simali$4.6$\mum$ in anions.
While the DFT-computed anharmonic spectra
provide fundamental information on the intrinsic
vibrational properties of molecules, 
only cascade simulations can capture
the emission processes that govern
astronomical observations.
Overall, these results highlight the necessity of
incorporating anharmonic cascade emission modeling
when interpreting astronomical spectra.

Observationally, while the identification of
cyanonaphthalenes through the $\simali$6--14$\mum$
complex could be complicated by the 6.2, 7.7, 8.6,
11.3 and 12.7$\mum$ AIB bands,
the detection of the C$\equiv$N stretching band
could allow one to infer, at least,
the presence of cyano-substituted PAHs
in space, particularly in the era of JWST
whose unique high sensitivity and
high resolution capabilities 
could potentially place the detection of
of this band on firm ground.

Admittedly, the detection of the C$\equiv$N stretching band 
does not necessarily pinpoint
the presence of cyanonaphthalenes,
as other cyano-substituted PAHs
may also emit at similar wavelengths.
Nevertheless, the computed emissivity
(in units of eV$\cm$) facilitates a direct
energy-budget analysis, allowing one to quantify
the precise fraction of the absorbed starlight energy
re-radiated into specific vibrational bands in the IR,
such as the C--H and C$\equiv$N stretches.
Such data are indispensable for calibrating
radiative cooling models and for interpreting
the integrated intensities from astronomical observations.
By providing the energy emitted per unit wavenumber,
our model cascade emission spectra enable a seamless
transition from theoretical molecular physics to observational
soectra of complex interstellar environments.

Let $\ICH$ and $\ICN$ (in units of eV) be the amount
of energy emitted in the C--H
and C$\equiv$N stretches
by a cyanonaphthalene molecule upon
absorption of a UV photon of energy $\EUV$.
As listed in Table~\ref{tab:integrated_energy_ratio},
we calculated $\ICH$ and $\ICN$ for
neutral, cationic and anionic 1- and 2-CNN
in four representative astrophysical environments
(e.g., in the Orion Bar or the M17 star-forming cloud,
upon absorption of a starlight photon of
$\EUV\approx9.8\eV$, 1-CNN cation would emit
$\simali$0.389$\eV$ in the nitrile stretching mode,
while only $\simali$0.034$\eV$ is deposited
into the C--H stretching).
%
Let $\PCN$ (in units of $\erg\s^{-1}\cm^{-2}\sr^{-1}$)
be the observed flux of the C$\equiv$N stretching band 
of an astronomical source.
By assuming the C$\equiv$N stretching band
exclusively arises from cyanonaphthalene, 
its column density can be derived from:
%
%
\begin{equation}
N\,({\rm cm}^{-2}) =
\left(\frac{4\pi}{1.602\times10^{-12}}\right)
\left(\frac{\tauabs}{\s}\right)
\left(\frac{\ICN}{\eV}\right)^{-1}
\left(\frac{\PCN}{\erg\s^{-1}\cm^{-2}\sr^{-1}}\right) ~~.
\end{equation}
Take the HII region of the Orion Bar---for which
$\PCN\approx9.80\times10^{-4}\erg\s^{-1}\cm^{-2}\sr^{-1}$
(see Yang \& Li 2025)---as an example.
If we assume neutral 1-CNN---for which
$\tauabs\approx732\s$ (under $U=10,000$)
and $\ICN\approx0.526\eV$
(see Table~\ref{tab:integrated_energy_ratio})---causes
the observed C$\equiv$N stretching band,
we obtain $N\approx 1.07\times10^{13}\cm^{-2}$.
Of course, as mentioned earlier, the observed
C$\equiv$N stretching band is unlikely produced
by a single molecule, 
this estimation would overestimate the abundance
of 1-CNN.\footnote{%
  McGuire et al.\ (2021) derived
$N\approx 7.35\times10^{11}\cm^{-2}$ for 1-CNN
and $N\approx 7.05\times10^{11}\cm^{-2}$ for 2-CNN
in the TMC-1 molecular cloud,
based on their rotational spectra.
}
Nevertheless, if the intrinsic strengths
of the C$\equiv$N band
of cyano-PAHs do not differ from
that of cyanonaphthalene by orders of
magnitudes, this derivation provides
an educational estimation of the amount
of all cyano-PAHs as a family in the ISM.
To be more accurate, we need to explore
a large number of cyano-substituted PAH
species and derive a mean $\ICN$ and $\tauabs$
and then compare with the observed $\PCN$.

\begin{figure*}
\centering
\includegraphics[width=\textwidth]{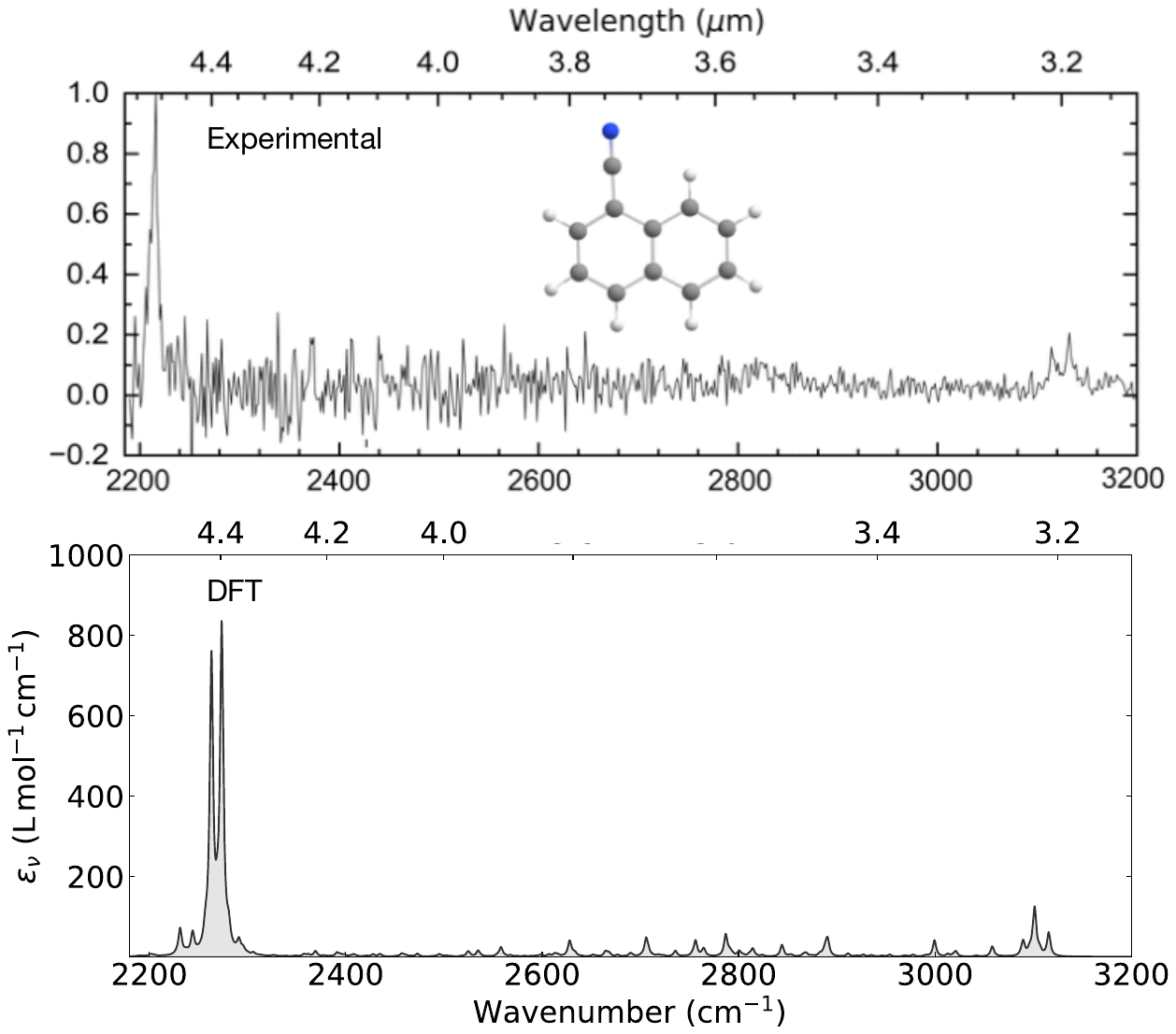}
\caption{Comparison between the experimental
cryogenic gas-phase IR spectrum of $1\text{-CNN}^+$
\citep{Palotas2024} 
and the theoretical spectrum calculated with VPT2
at the B3LYP/N07D level. 
The theoretical spectrum (bottom panel) was convoluted
with a Lorentzian profile of a full-width at half-maximum
(FWHM) of 4$\cm^{-1}$.
The MAE between the computed frequencies
and experimental benchmarks (see eq.\,\ref{eq:MAE})
is $\simali$47.8$\cm^{-1}$ for the C$\equiv$N
stretching region (i.e., the two prominent peaks
near 2200$\cm^{-1}$) and $\simali$27.2$\text{cm}^{-1}$
for the C--H stretching region
(i.e., the two peaks near 3100$\cm^{-1}$).
}
\label{fig:fig5}
\end{figure*}

Finally, we note that throughout the entire paper
we have taken the wavelength of the C$\equiv$N
stretch to be that computed from DFT
(i.e., $\simali$4.4$\mum$ for neutrals
and cations, and $\simali$4.6$\mum$
for anions). In general, the frequencies of
the C$\equiv$N stretching vibration fall
within the relatively narrow range of
$\simali$2210--2270$\cm^{-1}$,
corresponding to $\simali$4.41--4.52$\mum$
(see Wexler 1967).
To evaluate the accuracy of our DFT computations,
we compare the calculated anharmonic absorption
spectrum of $1\text{-CNN}^+$ and the experimental
cryogenic He-tagging IR photodissociation (IRPD)
spectrum \citep{Palotas2024}.
As shown in Figure~\ref{fig:fig5},
a line-by-line assignment reveals that
the MAE in peak positions\footnote{%
  The MAE is defined as
  \begin{equation}\label{eq:MAE}
  \text{MAE} = \frac{1}{2} \sum_{i=1}^{2}
   |\nu_{{\rm exp},i} - \nu_{{\rm DFT},i}| ~~,
 \end{equation}
 where$\nu_{{\rm exp},i}$ is
 the experimentally-measured frequency
 of the $i$-th vibrational band, and
 $\nu_{{\rm DFT},i}$ is the DFT-computed
 frequency of the $i$-th vibrational band.
}
is $\simali$47.8$\cm^{-1}$ for the C$\equiv$N stretches
(corresponding to the two prominent peaks
near 2200$\cm^{-1}$ or $\simali$4.5$\mum$
in Figure~\ref{fig:fig5})
and 27.2$\cm^{-1}$ for the aromatic C--H stretches
(corresponding to the two peaks near 3100$\cm^{-1}$
or 3.2$\mum$ in Figure~\ref{fig:fig5}).
If we invoke a scaling factor of $0.979$
for the nitrile stretching regions
(at $\nu \leq 2500\cm^{-1}$)
and $1.009$ for the C-H stretching region
(at $\nu > 2500\cm^{-1}$),
we would achieve a remarkably close
match to the experimental data.
On the other hand, Esposito et al.\ (2025)
have demonstrated that high-order DFT
compuations were able to improve the match
between the computational and experimental
data of cyanobenzene and 9-cyanoanthracene,
without invoking any scaling factors.
In searching for and identifying the IR signatures
of cyanonaphthalenes in space, we should keep
in mind that the actual C$\equiv$N band
may fall at $\simali$4.5$\mum$.

\section{Summary}
This study investigates the anharmonic vibrational
properties and cascade emission spectra of 1-CNN and 2-CNN
across multiple charge states under different astrophysical
environments, by integrating state-of-the-art anharmonic
calculations with a robust microcanonical sampling workflow.
Our major results are as follows:

\begin{enumerate}
\item We have computed the anharmonic vibrational spectra 
  of neutral, cationic and anionic 1-CNN and 2-CNN. It is 
  found that both the charge state and the substitution 
  pattern significantly influence the anharmonic 
  vibrational spectra of cyano-PAHs. In particular, 
  the strong and charge-sensitive C$\equiv$N stretching mode 
  provides a potentially powerful diagnostic feature for 
  identifying such species in space.

\item We have simulated the cascade emission of 
  neutral, cationic and anionic 1-CNN and 2-CNN in representative
  astrophysical environments. These cascade emission spectra could serve as a 
  high-fidelity diagnostic tool, as they encapsulate the 
  entire stochastic heating and cooling history of a molecule 
  following the absorption of a single photon. We have quantified the energy
dissipation efficiency through the C$\equiv$N stretch
feature relative to the total radiative cooling budget for these
species in different environments, and demonstrated how these
fractional emissivities, together with the observationally detected
 C$\equiv$N stretch flux, can be used to measure their column densities.
\end{enumerate}

\begin{acknowledgements}
We thank B.~Yang and the anonymous referee
for helpful comments.
KJL is supported in part by
NSFC~12503032.
The computations were enabled by resources provided 
by the National Academic Infrastructure for Supercomputing in Sweden (NAISS), 
partially funded by the Swedish Research Council through grant agreement 
no. 2022-06725.
\end{acknowledgements}

\bibliographystyle{aasjournal}

\begin{thebibliography}{}
\expandafter\ifx\csname natexlab\endcsname\relax\def\natexlab#1{#1}\fi

\bibitem[Andrews et al.(2015)]{Andrews2015}Andrews, H., Boersma, C., Werner, M.~W., et al.\ 2015, \apj, 807, 99

\bibitem[Allamandola et al.(1985)]{Allamandola1985}
Allamandola, L.~J., Tielens, A.~G.~G.~M., \& Barker, J.~R.\ 1985, \apjl, 290, L25

\bibitem[Allamandola et al.(1989)]{Allamandola1989}Allamandola, L.,
  Tielens, G. \& Barker, J.\ 1989, ApJS, 71, 733

\bibitem[Allamandola et al.(1999)]{Allamandola1999}Allamandola, L., Hudgins, D. \& Sandford, S. 1999, \apj, \textbf{511}, L115

\bibitem[Allamandola et al.(2021)]{Allamandola2021}Allamandola, L.,
  Boersma, C., Lee, T., Bregman, J., \& Temi, P.\ 2021, ApJL, 917, L35

\bibitem[Bakes et al.(2001)]{Bakes2001}Bakes, E., Tielens, A. \& Bauschlicher, C.\ 2001, \apj, \textbf{556}, 501


\bibitem[Barone et al.(2008)]{Barone2008}Barone, V., Cimino, P. \&
  Stendardo, E.\ 2008,  Journal of Chemical Theory \& Computation, 4, 751

\bibitem[Bauschlicher et al.(2009)]{Bauschlicher2009}Bauschlicher, C.,
  Peeters, E.. \& Allamandola, L.J.\ 2009, \apj, 697, 311

\bibitem[Bloino et al.(2012)]{Bloino2012}Bloino, J., Biczysko, M.,\&
  Barone, V.\ 2012,  Journal of Chemical Theory \& Computation, 8, 1015

\bibitem[Bloino et al.(2015)]{Bloino2015}Bloino, J., Biczysko, M. \&
  Barone, V.\ 2015, J. Phys. Chem. A, 119, 11862

\bibitem[Boersma et al.(2023)]{Boersma2023}Boersma, C., Allamandola,
  L.J., Esposito, V., et al.\ 2023, \apj, 959, 74

\bibitem[Burkhardt et al.(2021)]{Burkhardt2021}Burkhardt, A., Loomis,
  R., Shingledecker, C., Lee, K., Remijan, A., McCarthy, M., \&
  McGuire, B.\ 2021, Nature Astronomy, 5, 181

\bibitem[Cabezas et al.(2025)]{Cabezas2025}Cabezas, C., Ag{\'u}ndez, M., P{\'e}rez, C.,
                et al.\ 2025, \aap, 701, L8

\bibitem[Cernicharo et al.(2021)]{Cernicharo2021}Cernicharo, J.,
  Heras, A.M., Tielens, A.G.G.M., et al.\ 2001, ApJL, 546, L123

\bibitem[Cernicharo et al.(2024)]{Cernicharo2024}Cernicharo, J.,
  Cabezas, C., Fuentetaja, R., et al.\ 2024, \aap, 690, L13

\bibitem[Chen(2018)]{Chen2018apjs}Chen, T.\ 2018, ApJS, 238, 18

\bibitem[Chen et al.(2018)]{Chen2018aa}Chen, T., Mackie, C., Candian,
  A., Lee, T., \& Tielens, A.G.G.M.\ 2018, \aap, 618, A49

\bibitem[De Frees et al.(1993)]{DeFrees1993}Frees, D., Miller, M.,
  Talbi, D., Pauzat, F.,\& Ellinger, Y.\ 1993, ApJ, 408, 530

\bibitem[Dixon et al.(2015)]{Dixon2015}Dixon, A., Khuseynov, D. \& Sanov, A.\ 2015, \jcp, 143, 134306

\bibitem[{Draine \& Li(2001)}]{Draine2001}Draine, B.~T., \& Li, A.\ 2001, ApJ, 551, 807

\bibitem[Esposito et al.(2024a)]{Esposito2024a}Esposito, V.,
  Fortenberry, R., Boersma, C., Maragkoudakis, A., \& Allamandola,
  L.J.\ 2024, MNRAS, 531, L87

\bibitem[Esposito et al.(2024b)]{Esposito2024b}Esposito, V.,
  Fortenberry, R., Boersma, C., \& Allamandola, L.J.\ 2024,  ACS Earth
  \& Space Chemistry, 8, 1890

\bibitem[Esposito et al.(2025)]{Esposito2025}Esposito, V., Ferrari,
  P., Palmer, C., et al.\ 2025, J. Phys. Chem. Lett., 2025, 16, 1296

\bibitem[Frisch et al.(2016)]{g16}Frisch, M.~J., et al.\ 2016, Gaussian~16 Revision C.01

\bibitem[Gulania et al.(2020)]{Gulania2020}Gulania, S., Jagau, T., Sanov, A. \& Krylov, A.\ 2020, {\em Physical Chemistry Chemical Physics}, 22, 5002

\bibitem[Hudgins \& Allamandola(2005)]{Hudgins2005}Hudgins, D.~M., \&
  Allamandola, L.~J.\ 2005, Steps toward Identifying PAHs: A Summary
  of Some Recent Results, in {\it Astrochemistry: Recent Successes and
    Current Challenges}, 231, 443

\bibitem[Jiang et al.(2026)]{Jiang2026} Jiang, X.,  Fang, T., \&
  Yan, S.\ 2026, Sci. China-Phys. Mech. Astron, 69, 279512 (DOI: 10.1007/s11433-025-2965-5)
  
\bibitem[Kirnosov et al.(2017)]{Kirnosov2017}Kirnosov, N. \& Adamowicz, L.\ 2017, {\em Chemical Physics Letters}, 676, 32

\bibitem[Kwon et al.(2003)]{Kwon2003}Kwon, C., Kim, H., \& Kim, M.\ 2003, J. Phys. Chem. A, 107, 10969

\bibitem[Langhoff (1996)]{Langhoff1996}Langhoff, S.\ 1996, J. Phys. Chem., 100, 2819
  
\bibitem[Lee et al.(1988)]{Lee1988}Lee, C., Yang, W., \& Parr, R.~G.\ 1988, \prb, 37, 785

\bibitem[L{\'e}ger \& Puget(1984)]{Leger1984}
L{\'e}ger, A., \& Puget, J.~L.\ 1984, \aap, 137, L5

\bibitem[Li(2020)]{Li2020}Li, A.\ 2020, Nature Astronomy, 4, 339

\bibitem[Li et al.(2024a)]{Li2024a}Li, K.J., Li, A., Yang, X.J., \&
  Fang, T.\ 2024a, ApJ, 961, 107

\bibitem[Li et al.(2024b)]{Li2024b}Li, K.J., Li, A., Yang, X.J., \& Fang, T.\ 2024b, MNRAS, 529, 4425

\bibitem[Liu et al.(2022)]{Liu2022}Liu, J., Feng, R., Zhou, L., Gai,
  F., \& Zhang, W.\ 2022, J. Phys. Chem. Lett., 2022, 13, 9745

\bibitem[Mackie et al.(2015)]{Mackie2015}Mackie, C., Candian, A.,
  Huang, X., et al.\ 2015, \jcp, 143, 134302

\bibitem[Mackie et al.(2016)]{Mackie2016}Mackie, C., Candian, A.,
  Huang, X., et al.\ 2016, \jcp, 145, 084313

\bibitem[Mackie et al.(2018)]{Mackie2018}Mackie, C., Chen, T.,
  Candian, A., Lee, T. \& Tielens, A.G.G.M.\ 2018, \jcp, 149, 134302

\bibitem[McGuire et al.(2018)]{McGuire2018}McGuire, B., Burkhardt, A.,
  Kalenskii, S., Shingledecker, C., Remijan, A., Herbst, E. \&
  McCarthy, M.\ 2018, Science, 359, 202

\bibitem[McGuire et al.(2021)]{McGuire2021}McGuire, B., Loomis, R.,
  Burkhardt, A., et al.\ 2021, Science, 371, 1265

\bibitem[Palotás et al.(2024)]{Palotas2024}Palot\'as, J., Daly, F.,
  Douglas-Walker, T. \& Campbell, E.\ 2024, Physical Chemistry
  Chemical Physics, 26, 4111

\bibitem[Rajasekhar et al.(2022)]{Rajasekhar2022}Rajasekhar, B.,
  Dharmarpu, V., Das, A., Shastri, A., Veeraiah, A., \& Krishnakumar,
  S.\ 2022, JQSRT, 283, 108159

\bibitem[Ricca et al.(2012)]{Ricca2012}Ricca, A., Bauschlicher, C.,
  Boersma, C., Tielens, A.G.G.M., \& Allamandola, L.J.\ 2012, \apj,
  754, 75

\bibitem[Sita et al.(2022)]{Sita2022}Sita, M., Changala, P., Xue, C.,
  et al.\ 2022, ApJL, 938, L12

\bibitem[Stockett et al.(2025)]{Stockett2025}Stockett, M., Esposito,
  V., Ashworth, E., Jacovella, U., \& Bull, J.\ 2025, ACS Earth \&
  Space Chemistry, 9, 382

\bibitem[Tielens(2008)]{Tielens2008}Tielens, A.G.G.M.\ 2008, ARA\&A, 46, 289
  
\bibitem[Wang \& Landau (2001)]{Wang2001}Wang, F., \& Landau, D.\
  2001, PRL, 86, 2050

\bibitem[Wenzel et al.(2024a)]{Wenzel2024a}Wenzel, G., Speak, T.,
  Changala, P., et al.\ 2024, Nature Astronomy, 9, 262

\bibitem[Wenzel et al.(2024b)]{Wenzel2024b}Wenzel, G., Cooke, I.,
  Changala, P., et al.\ 2024, Science, 386, 810

\bibitem[Wenzel et al.(2025)]{Wenzel2025}Wenzel, G., Gong, S., Xue,
  C., et al.\ 2025, ApJL, 984, L36

\bibitem[Wexler(1967)]{Wexler1967}Wexler, A.\ 1967, Applied
  Spectroscopy Reviews, 1, 29. doi:10.1080/05704926708547581

  
\bibitem[Xu et al.(2024a)]{Xu2024a}Xu, R., Jiang, Z., Yang, Q.,
  Bloino, J., \& Biczysko, M.\ 2024,  Journal of Computational
  Chemistry, 45, 1846

\bibitem[Xu et al.(2024b)]{Xu2024b}Xu, Y., \& Biczysko, M.\ 2024,
  Frontiers in Chemistry, 12, 1439194

\bibitem[Zhang et al.(2026)]{Zhang2026} Zhang, C., Liu, T., Juvela,
    M., et al.\ 2026,  Sci. China-Phys. Mech. Astron, 69, 289512 (DOI:
    10.1007/s11433-026-2964-7)
  
\end{thebibliography}

\begin{onecolumn}{}
\begin{table}[p]
\footnotesize
\caption{Absorption and emission properties of neutral, cationic
  and anionic 1- and 2-CNN in four representative astrophysical
  environments.}
\label{tab:integrated_energy_ratio}
\setlength{\tabcolsep}{4.5pt}
\begin{tabular}{lcccccccccccccccr}
\toprule
& \multicolumn{4}{c}{\textbf{Diffuse ISM}} & \multicolumn{4}{c}{\textbf{Red Rectangle}} & \multicolumn{4}{c}{\textbf{NGC 2023}} & \multicolumn{4}{c}{\textbf{Orion Bar, M17}} \\
& \multicolumn{4}{c}{\textbf{(MMP83, $U$\,=\,1)}}
  & \multicolumn{4}{c}{\textbf{($\Teff$\,=\,8,000$\K$,
    $U$\,=\,10$^5$)}} & \multicolumn{4}{c}{\textbf{($\Teff$\,=\,22,000$\K$, $U$\,=\,10$^3$)}} & \multicolumn{4}{c}{\textbf{($\Teff$\,=\,40,000$\K$, $U$\,=\,10$^4$)}}\\
\cmidrule(lr){2-5} \cmidrule(lr){6-9} \cmidrule(lr){10-13} \cmidrule(lr){14-17}
Molecule & $\EUV^a$ & $\tauabs^b$ & $\ICN^c$ & $\ICH^d$ & $\EUV^a$ & $\tauabs^b$ & $\ICN^c$ & $\ICH^d$ & $\EUV^a$ & $\tauabs^b$ & $\ICN^c$ & $\ICH^d$ &$\EUV^a$ & $\tauabs^b$ & $\ICN^c$ & $\ICH^d$ \\
 & (eV) & (s) & (eV) & (eV) & (eV) & (s) & (eV) & (eV) & (eV) & (s) & (eV) & (eV) & (eV) & (s) & (eV) & (eV)\\
\midrule
1-CNN$^{+}$ & 8.6 & $3.7 \times 10^7$ & 0.353 & 0.032 & 5.7 & 816 & 0.253 & 0.028 & 9.3 & 8213  & 0.376 & 0.033 & 9.8 & 732 & 0.389 & 0.034\\
1-CNN$^{0}$ & 8.6 & $3.7 \times 10^7$ & 0.473 & 0.290 & 5.7 & 816 & 0.331 & 0.234 & 9.3 & 8213 & 0.506 & 0.300 & 9.7 & 732 & 0.526 & 0.305 \\
1-CNN$^{-}$ & 8.6 & $3.7 \times 10^7$ & 1.499 & 0.208 & 5.7 & 816 & 1.035 & 0.177& 9.3 & 8213 & 1.608 & 0.212 & 9.8 & 732 & 1.669 & 0.214 \\
\midrule
2-CNN$^{+}$ & 8.6 & $3.7 \times 10^7$ & 0.173 & 0.025 & 5.7 & 816 & 0.130 & 0.021 & 9.3 & 8213 & 0.183 & 0.025 & 9.8 & 732 & 0.188 & 0.026\\
2-CNN$^{0}$ & 8.6 & $3.7 \times 10^7$ & 0.562 & 0.210 & 5.7 & 816 & 0.399 & 0.174 & 9.3 & 8213 & 0.601 & 0.216 & 9.7 & 732 & 0.624 & 0.219 \\
2-CNN$^{-}$ & 8.6 & $3.7 \times 10^7$ & 1.289 & 0.130 & 5.7 & 816 & 0.918 & 0.112 & 9.3 & 8213 & 1.373 & 0.133 & 9.8 & 732 & 1.421 & 0.135 \\
\bottomrule
\addlinespace[0.5ex]
\end{tabular}
\\
$^a$ Mean energy of the photons absorbed by the
  molecule. As we do not distinguish the UV absorption properties
  of 1- and 2-CNN and their charge states, $\EUV$ is the same for all
  species in a given astrophysical environment. We note that $\EUV$ is
  independent of $U$.\\
 $^b$ Photon absorption time scale. Again, as we
  do not distinguish the UV absorption properties of 1- and 2-CNN and
  their charge states, $\tauabs$ is the same for all species in a
  given astrophysical environment. We note that $\tauabs$ is inversely
  proportional to $U$.\\
$^c$ Amount of energy emitted in the
  C$\equiv$N stretch band upon absorption of a starlight photon of energy
  $\EUV$. We note that $\ICN$ is independent of $U$.\\
$^d$ Amount of energy emitted in the 3.3$\mum$
  C--H band upon absorption of a starlight photon of energy
  $\EUV$. We note that $\ICH$ is independent of $U$.
\end{table}
\end{onecolumn}{}

\appendix
\section{Appendix}
\begin{figure*}
\centering
\includegraphics[width=\textwidth]{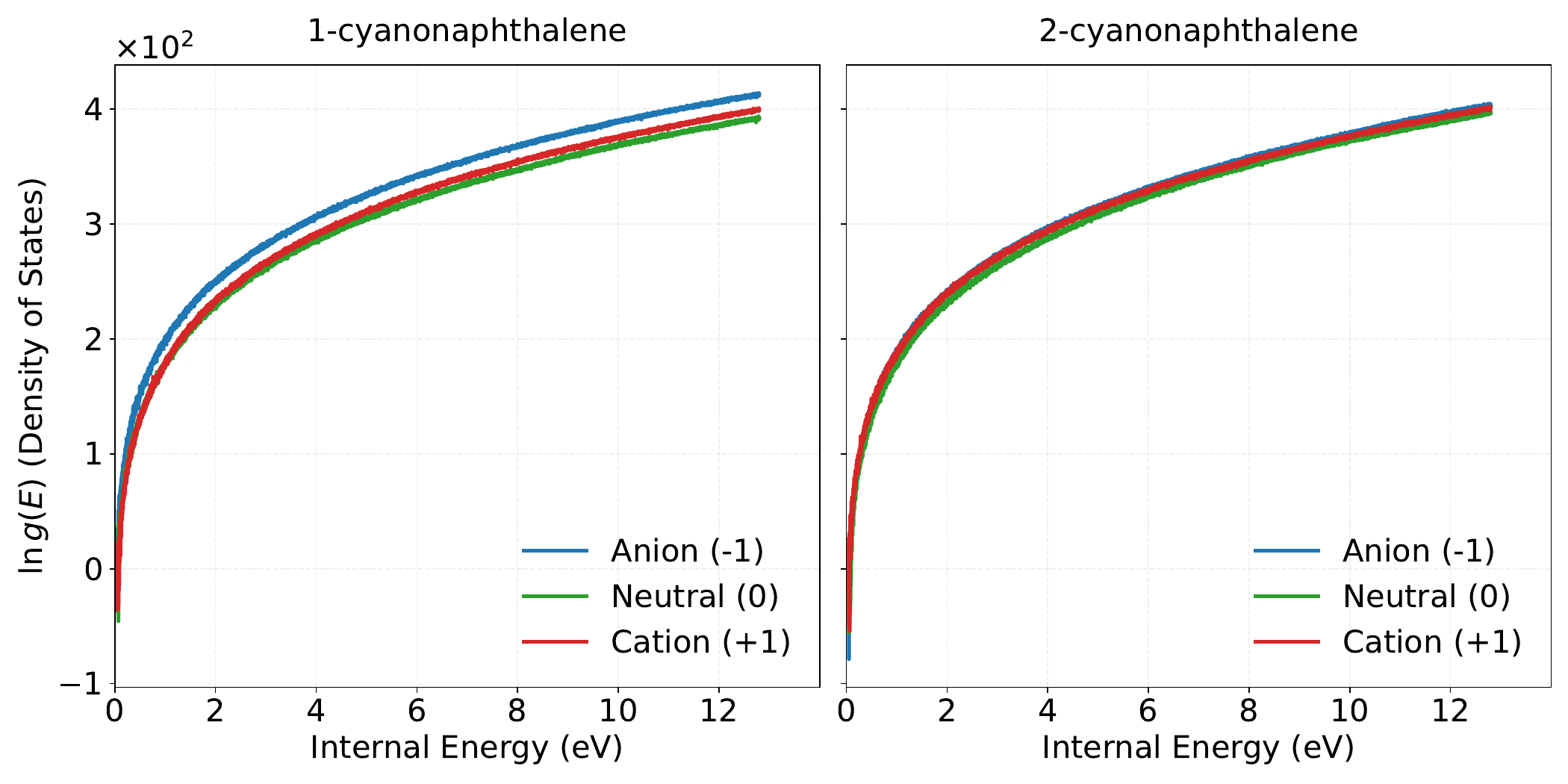}
\caption{Vibrational DOS, expressed as $\ln g(E)$,
for 1-CNN (left panel) and 2-CNN (right panel)
in three different charge states:
anion (blue), neutral (green), and cation (red),
obtained using the WL sampling method based
on the B3LYP/N07D+VPT2 anharmonic potential.
The internal energy $E$ ranges from $0$ to $12\eV$.
The overlapping curves for both isomers demonstrate
the consistent thermodynamic behavior of the CNN species,
while the subtle differences between charge states
reflect the variations in vibrational anharmonicity
and mode coupling upon electron addition or removal.
}
\label{fig:figapp1}
\end{figure*}

Figure \ref{fig:figapp1} shows the calculated DOS
for 1-CNN and 2-CNN across three charge states.
They all exhibit a near-exponential growth
with internal energy, consistent with the behavior of
polycyclic aromatic species. For both isomers,
the cation and anion states show slightly higher
densities of states compared to the neutral species
at high internal energies. This divergence is attributed
to the modification of the vibrational potential energy
surface upon ionization, which affects the coupling
constants $\chi_{ij}$ and the fundamental frequencies
$\nu_i$. The smooth and converged profiles of
$\ln g(E)$ validate the robustness of the WL sampling,
ensuring that the subsequent cascade spectra are based
on a statistically sound representation
of the vibrational phase space.

\end{document}